\documentstyle[preprint,aps,eqsecnum]{revtex}
\def\mycomm#1{\hfill\break\strut\kern-3em{\tt ====> #1}\hfill\break}
\def\mycommNL#1{\strut\kern-3em{\tt ====> #1}\hfill\break}

\def\sbar{\hbox{$\bar s$}}
\def\cbar{\hbox{$\bar c$}}

\def\thetac{\hbox{$\Theta_c$}}
\def\thetab{\hbox{$\Theta_b^+$}}
\def\thetap{\hbox{$\Theta^+$}}

\def\eqref#1{(\ref{#1})}

\makeatletter
\def\hlinewd#1{\noalign{\ifnum0=`}\fi
\hrule \@height #1 \futurelet \reserved@a\@xhline}
\def\hwhiteline{\noalign
{\ifnum0=`}\fi\hrule 
\@height 0pt\vskip 1.0ex\futurelet \reserved@a\@xhline}
\makeatother
\def\gray{\special{ps: 0.40 setgray}}
\def\black{\special{ps: 0.0 setgray}}

\newcommand{\mydraft}{
\newcount\timecount
\newcount\hours \newcount\minutes  \newcount\temp \newcount\pmhours

\hours = \time
\divide\hours by 60
\temp = \hours
\multiply\temp by 60
\minutes = \time
\advance\minutes by -\temp
\def\hour{\the\hours}
\def\minute{\ifnum\minutes<10 0\the\minutes
    \else\the\minutes\fi}
\def\clock{
\ifnum\hours=0 12:\minute\ AM
\else\ifnum\hours<12 \hour:\minute\ AM
\else\ifnum\hours=12 12:\minute\ PM
    \else\ifnum\hours>12
	 \pmhours=\hours
	 \advance\pmhours by -12
	 \the\pmhours:\minute\ PM
	 \fi
    \fi
\fi
\fi
}
\def\fullclock{\hour:\minute}
\begin{centering}
\gray
\font\Hugett  =cmtt12 scaled\magstep4
\hbox{\Hugett Draft:\today,\clock}
\black
\end{centering}
\vskip -1.7cm
$\phantom{a}$
} 

\def\beq#1{\begin{equation} \label{#1}}
\def\eeq{\end{equation}}

\def\ket#1{\left\vert #1\right\rangle}

\newskip\humongous \humongous=0pt plus 1000pt minus 1000pt

\newif\ifdtup

\begin{document}
{\tighten
\preprint
{\vbox{\hbox{\today}
\hbox{}
\hbox{TAUP 2744-03}
\hbox{WIS/21/03-JULY-DPP}
\hbox{hep-ph/0307343} }}

\title{The anticharmed exotic baryon $\Theta_c$ and its relatives}

\author{Marek Karliner\,$^a$\thanks{e-mail: \tt marek@proton.tau.ac.il}
\\
and
\\
Harry J. Lipkin\,$^{a,b}$\thanks{e-mail: \tt ftlipkin@clever.weizmann.ac.il}
}
\address{ \vbox{\vskip 0.truecm}
$^a\;$School of Physics and Astronomy \\
Raymond and Beverly Sackler Faculty of Exact Sciences \\
Tel Aviv University, Tel Aviv, Israel\\
\vbox{\vskip 0.0truecm}
$^b\;$Department of Particle Physics \\
Weizmann Institute of Science, Rehovot 76100, Israel \\ 
and\\
High Energy Physics Division, Argonne National Laboratory \\
Argonne, IL 60439-4815, USA\\
}
\maketitle

\begin{abstract}%
Motivated by the recent discovery of the exotic $S={+}1$ narrow
baryon resonance $\Theta^+$ at 1540 MeV,
with quark content $uud\,d\sbar$,
we conjecture the existence of its anti-charmed analogue \thetac
with quark content $uudd\cbar$,
and compute its likely properties. We rely on the recently constructed 
model of a novel kind of a pentaquark with an unusual color structure which
provides a good approximation to the \thetap\ mass. We expect that 
\thetac\ is an isosinglet with $J^P={1\over2}^+$ and estimate its mass at 
$2985 \pm 50$ MeV. We also discuss another possible exotic baryon resonance
containing heavy quarks, the \thetab, a $uudd\bar b$ state, and estimate
$m_{\thetab}=6398\pm 50$ MeV.
These states should appear as unexpectedly narrow peaks in 
$D^-p$, $\bar D^0 n$,
$B^0 p$ and $B^+n$ mass distributions.

\end{abstract}%
} 

\pagebreak
\section{Introduction }

The recent experimental discovery of an exotic
exotic 5-quark $K N$ resonance \cite{Kyoto,Russia,Stepanyan:2003qr}, 
\thetap\
with $S={+}1$, a mass of 1540 MeV and a very small width $\sim$20 MeV
opens up the possibility that other similar exotic baryon 
resonances might exist. 

The simplest quark configuration with the 
quantum numbers of the \thetap\ is $uudd\sbar$, with the light $u$ and $d$
quark likely coupled to an isosinglet.
In a recent paper \cite{NewPenta}
we proposed  an interpretation of the \thetap\ as $uudd\sbar$
pentaquark consisting of an $I=0$ color antitriplet $ud$ diquark 
coupled to an $I=0$ color triplet $uds$ triquark, with one unit of 
relative orbital angular momentum between the two clusters.
This unusual color structure turns out to minimize the hyperfine
interaction of the color magnetic moments of the constituents.
The resulting tri-diquark has $I=0$ and $J^P=1/2^+$, 
in accordance with expectations based on the Skyrme model
\cite{SMKNa,SMKNb}.
A rough estimate of the mass in \cite{NewPenta} yields
1592 MeV, about 3\% off the experimental value.
       
There is nothing in QCD that prevents the existence of states with more
than 3 quarks, or mesons with additional quarks, on top of a quark and an
antiquark. Yet, the spectacular phenomenological success of the quark
model based on 3-quark baryons and quark-antiquark mesons made it almost a
dogma that other, ``exotic" states do not exist. Now, that the \thetap\ has
been discovered experimentally, it is clear that this dogma needs a deep
revision.

Any five-quark explanation of the \thetap\ should have the four quarks $uudd$
coupled to isospin zero and an additional heavier antiquark. This immediately
raises the question of the flavor of the antiquark. If any $(uudd\bar s)$ model
is good for a strange antiquark, why not also a similar $(uudd\bar Q)$ model
with any heavy antiquark Q like  charmed or bottom? QCD tells us that the only
difference QCD sees between these different flavors is their mass. So it is
reasonable to look for a narrow resonance also in $D^- p$, $\bar D^0 n$ or $B^0 p$,
$B^+ n$.  These should appear as unexpectedly narrow peaks in mass
distributions.

In this work we focus on the possibility that such exotic baryons do exist,
and compute their properties, using 
the model constructed in Ref.~\cite{NewPenta}.

\section{The diquark-triquark model}
Most quark model treatments of multiquark spectroscopy use the color-magnetic 
short-range hyperfine interaction\cite{DGG} as the dominant mechanism for
possible binding. The  application of this interaction by Jaffe\cite{Jaffe} to
treat the exotic color configurations not found in normal hadrons used a
color-spin $SU(6)$ algebra in which the  hyperfine interaction between two quarks
denoted by $i$ and $j$ is written as 

\beq{vhypI}
V_{hyp} =-V(\vec \lambda_i \cdot \vec \lambda_j)(\vec \sigma_i \cdot \vec \sigma_j)
\end{equation} 
where $\vec \lambda$ and $\vec \sigma$ denote the generators of $SU(3)_c$ and
the Pauli spin operators, respectively. 
Jaffe has used the sign and magnitude of the
$\Delta$-nucleon mass splitting as input for the sign and strength of the
hyperfine interaction. 
The quark-quark interaction \eqref{vhypI} 
is seen to be attractive
in states symmetric in color and spin where 
$(\vec \lambda_i \cdot \vec \lambda_j)$ and $(\vec \sigma_i \cdot \vec
\sigma_j)$ have the same sign 
and repulsive in antisymmetric states whee they have opposite signs.
This then leads to the "flavor-antisymmetry" principle\cite{Lipflasy}: the
Pauli principle forces two identical fermions at short distances to be in a
state that is antisymmetric in spin and color where the hyperfine interaction
is repulsive.  Thus the hyperfine interaction is always repulsive between two
quarks of the same flavor, such as the like-flavor $uu$ and $dd$  pairs in the
nucleon or pentaquark. 

This flavor antisymmetry suggests that the bag or single-cluster models
commonly used to treat normal hadrons may not be adequate for multiquark
systems. In such a state with identical pair correlations for all pairs in the
system all same-flavor quark pairs are necessarily in a higher-energy
configuration due to the repulsive nature of their hyperfine interaction, The
$uudd\bar s$ pentaquark is really a complicated five-body system  where the
optimum wave function to give minimum color-magnetic energy can require
flavor-dependent spatial pair correlations for different pairs in the system;
e.g. that keep the like-flavor $uu$ and $dd$ pairs apart,  while minimizing the
distance and optimizing the color couplings within the  other pairs. This is
the physical basis for the success of the diquark-triquark model
\cite{NewPenta}.  An even more complicated spatial configuration might well do
better.

To see a simple picture of the model constructed in  Ref.~\cite{NewPenta} take
a $K^+$ and a neutron and put them just  far enough apart so that they are out
of the range of the very short  range Fermi color hyperfine interaction.  Now
take one of the $d$-quarks in the neutron and move it over  to the kaon and
recouple the color and spin to optimize the hyperfine  interaction. 

Moving the quark a distance from a point $r_1$ to a point $r_2$ while doing
nothing to the kaon requires an energy in the potential model of the neutron of
$V(r_2) - V(r_1)$, where $V$ is the confining potential;  e.g. Coulomb +
linear.  The energy is in the color electric field that has been created
between $r_1$ and $r_2$.  This distance costs color electric energy. Recoupling
the color and spins of the triquark gains hyperfine energy. But because the
triquark is a color-electric triplet like the  quark, the recoupling does not
change the color electric field in the approximation where the spatial
extension of the triquark is neglected.

The change in color-electric energy can also be described in the approximation
of a point triquark and point diquark by solving the Schroedinger equation for 
a quark-antiquark pair in the confinement potential.
One has to balance the gain in hyperfine energy with the excitation energy of
the relative diquark-triquark system in the confinement potential. The rough 
calculation in Ref.~\cite{NewPenta} suggests that this tradeoff between 
the hyperfine and the confining interaction reproduces the measured mass of
the \thetap.
 
In this picture we can replace the strange 
antiquark by a charmed antiquark and get a narrow resonance in $D^-$-proton 
scattering. 
This might be seen by FOCUS by checking the invariant 
mass distribution of $D^-$-proton, to see whether there may be a narrow 
peak like the \thetap. This possibility is discussed in detail in the next
section.

\section{Exotic baryons with 
\,\MakeLowercase{$\bar c$} 
\,or 
\,\MakeLowercase{$\bar b$} 
\,instead of
\,\MakeLowercase{$\bar s$} }
The detailed Skyrme model prediction for the anti-strange pentaquark
\cite{SMKNb} has now been 
beautifully confirmed by the data \cite{Kyoto,Russia,Stepanyan:2003qr}.
However, the exotic baryons in which the \sbar\ is replaced 
by a heavier antiquark
cannot be treated within the Skyrme model approach 
of Ref.~\cite{SMKNb}, in which the
strange antiquark is in the same $SU(3)_f$ multiplet as $u$ and $d$,
and symmetry breaking by quark mass differences is treated
as a first-order perturbation. This approach may be fine for the $s$
quarks, but it not valid for heavier quarks such as $c$ and $b$.

In contrast,
the explicit model in Ref.~\cite{NewPenta} is easily generalized to
include any mass for the antiquark. This mass appears only in the
hyperfine interaction between the antiquark and the $ud$ pair in the
triquark, with a coefficient inversely proportional to the
antiquark mass. All that is necessary to consider an antiquark of any
flavor is to adjust this coefficient to the appropriate inverse quark mass.
   
Ref.~\cite{NewPenta} treats the hyperfine interactions in the
diquark-triquark model in the $SU(3)$ limit; i.e. with the mass of the strange 
antiquark equal to the masses of the $u$ and $d$  quark. The necessary 
generalization to include any antiquark flavor and mass is easily 
incorporated by
separating the quark-quark and quark-antiquark 
interaction and changing the coefficient of the quark-antiquark interaction 
to correspond to the correct antiquark mass. 

To show explicitly how the results of Ref.~\cite{NewPenta} can be generalized,
we first review in detail the calculation leading to their result.

We use the general form of the color-spin hyperfine interaction
\cite{Jaffe}
for systems containing both quarks and antiquarks:

\beq{jafhyp}
  V = (v/2)[\bar C(tot) - 2\bar C(Q) - 2\bar C(\bar Q) + 16N]
 \end{equation} 
where $V$ is the total hyperfine contribution to 
the mass of the system,
and
$v$ is a parameter defining the strength of the interaction, normalized
by computing the value of $V$ for the nucleon and the $\Delta$,
and equating it with the experimental value of the $\Delta$-nucleon 
mass splitting,
\beq{vhyp}
 M(\Delta)-M(N) 
 = V(\Delta) - V(N) = 16v, 
 \end{equation} 
 $\bar C(tot)$,
$\bar C(Q)$ and $\bar C(\bar Q)$ denote respectively the values  for the whole
system and for the subsystems of all the quarks and all the antiquarks in the
system of the following linear combination of Casimir operators
\beq{casbar}
\bar C = C_6 - C_3 -(8/3)S(S+1)     
\end{equation}
 $C_6$ and $C_3$ denote the eigenvalues of the Casimir  operators of the
$SU(6)$ color-spin and $SU(3)$ color groups respectively,  and 
$S$ and $N$ denote the total spin and the number of quarks in the system. 

The diquark, triquark and meson states are labeled
in the conventional notation $\ket{D_6,D_3,S,N}$
\cite{Patera,Sorba}
where $D_6$ and $D_3$ denote the dimensions of the color-spin $SU(6)$ and
color SU(3) representations in which the multiquark states are classified,
 .
\beq{diq1}
\ket{{\rm diquark}(S=1)} =   \ket{(2q)_{21}^1} = \ket{\bar {21},6,1,2}   
\end{equation}
\beq{diq0}
\ket{{\rm diquark}(S=0)}  = \ket{(2q)_{21}^0} =  \ket{\bar {21},\bar 3,0,2}  
\end{equation}
\beq{triq}
\ket{{\rm triquark}(S=1/2)} = \ket{(2q\bar s)_{21}^{(1/2)}} = \ket{6,3,1/2,3}
\end{equation}
\beq{meson}  
\ket{{\rm meson}}= \ket{(q\bar s)_{1}^{0}} = \ket{1,1,0,2}
\end{equation} 

Then
\begin{equation}
\bar C(2q)_{21}^1 =[(160/3) - (40/3) -(16/3)] = (104/3)
\end{equation}
\begin{equation}
\bar C(q) = \bar C(\bar s) = [(70/3) - (16/3) - 2 ] = 16
\end{equation}
\begin{equation}
\bar C(q \bar s)_1 =[(0) ]= 0 
\end{equation}

The interaction is easily evaluated for the diquark states
(\ref{diq1}-\ref{diq0}) by
substituting the eigenvalues of the Casimir operators
\cite{Patera,Sorba}:
\begin{eqnarray}
  C_6(6) &= (70/3)  &\\
  C_6(21) &= (160/3) & \\
   C_3(3) &=(16/3)  &\\
  C_3(6) &=(40/3)  &
\end{eqnarray}
We then obtain

\beq{hfdiq1}
 V(2q)_{21}^1 = -(v/2)[(160/3) - (40/3) -(16/3) - 32] = -(8/3)(v/2) 
 \end{equation}
\beq{hfdiq2}
V(2q)_{21}^0 = -(v/2)[(160/3) - (16/3) - 32] = -(16)(v/2) 
 \end{equation}

For the triquark and meson states we obtain    

\beq{hftriq}
 V(2q\bar s)_{21}^1 = (v/2)[16 - 2(104/3) - 32 + 48] = -(112/3)(v/2)
 \end{equation}
\beq{hfmes}
V(q\bar s)_0 = (v/2)[ - 64 + 32] = (-32)(v/2)
\end{equation}
We now separate the contributions to the triquark hyperfine
  interaction(\ref{hftriq}) into the quark-quark and quark-antiquark
  contributions by noting that the hyperfine interaction in the diquark that is 
  in the triquark
 is $-(8/3)(v/2)$ , We can then write the generalized triquark hyperfine 
 interaction for the case where the antiquark mass is  $m_Q$., 
\beq{hftriqsep}
 V(2q\bar s)_{21}^1 = -(8/3)(v/2) - \zeta \cdot (104/3)(v/2)
 \end{equation}
where $\zeta = m_u/m_Q$. For the generalized meson,
\beq{hfmessep}
V(q\bar s)_0 = - \zeta \cdot (32)(v/2)
\end{equation}

The hyperfine interaction in the diquark of our diquark-triquark model
is equal to the hyperfine interaction in the nucleon. 
$$V(N) = V(\Lambda) =  V(2q)_{21}^0  = -8v $$

Thus the difference between the
hyperfine interactions in the diquark-triquark system will differ from that in
the kaon-nucleon system only by the difference between the triquark and the
kaon.

\beq{WX11a}
V(2q\bar s)_{21}^1 - V(q\bar s)_0 =  -(1+\zeta)\cdot (4/3)v 
\end{equation}
\beq{WX11b}
[V(2q\bar s)_{21}^1 + V(d_{21}^0)] - [V(q\bar s)_0 + V(N)]=
-{{1+\zeta}\over{12}}\cdot [M(\Delta)-M(N)]
\end{equation}
The hyperfine interaction is greater by $(1+\zeta)\cdot (1/12)[M(\Delta)-M(N)]$
for the diquark-triquark system than for the kaon nucleon system.
This gives previous result\cite{NewPenta} of $-(8/3)v =-(1/6)(M_\Delta - 
M_N)$ for
$\zeta=1$ .
This result is obtained without any flavor symmetry assumption and holds for
any mass antiquark.

\subsection{The anticharmed exotic baryon $\thetac=uudd\bar c$}
We now apply this formalism to the specific case of the
anticharmed exotic baryon \thetac, i.e. the $uudd\bar c$ pentaquark.
We use effective quark masses that fit the low-lying mass spectrum
\cite{NewPenta}:
\beq{qmass}
m_u =  m_d = 360
\hbox{\ MeV};\quad
m_s=  540
\hbox{\ MeV};\quad
m_c= 1710
\hbox{\ MeV};\quad
m_b = 5050 \hbox{\ MeV}\,.
\label{quarkmasses}
\end{equation}
from this we find a very rough estimate of the $ud$ diquark and $ud\bar c$
triquark effective masses
\beq{qmass2}
m_{ud} =  720
\hbox{\ MeV};\qquad
m_{ud\bar c}= 2430 \hbox{\ MeV}\,,
\end{equation}
so that the reduced mass for the relative motion of the 
$ud$ diquark and $ud\bar c$ triquark system is 
$m_r(\{ud\}\hbox{-}\{ud\bar c\})= 555$ MeV. This reduced mass is fairly
close to the reduced mass of the $c\bar s$ system used to describe the
internal structure of the $D_s$ spectrum, $m_r(c\bar s)=  410$ MeV.
The dependence of the excitation energy on the reduced mass is expected to
be rather small, e.g. the $\psi^\prime-J/\psi$ splitting is 589 MeV
vs.  563 MeV for $\Upsilon^\prime-\Upsilon$.
Using the proximity of reduced masses,
we can obtain a rough estimate of the $P$-wave excitation energy
in the $\{ud\}\hbox{-}\{ud\bar c\}$ diquark-triquark system
\cite{NewPenta}, using 
the relevant experimental information about the
$D_s$ system,
\cite{Aubert:2003fg,CLEO,BELLE-Ds,RPP}
\begin{equation}
\delta E^{P-wave} \approx  207\ \hbox{MeV}\,.
\label{EPwave}
\end{equation}
From eq.~(\ref{WX11b}) we infer that without the $P$-wave excitation energy 
the $\{ud\}\hbox{-}\{ud\bar c\}$ diquark-triquark mass is 
\begin{equation}
m^0_{\{ud\hbox{-}ud\bar c\}}= m_N + m_D 
-{1\over 12}\,(1+\zeta_c)\, \left[M(\Delta)-M(N)\right] \approx 2778 \ \hbox{MeV}\,.
\label{M0uuddcbar}
\end{equation}
where $\zeta_c =  m_u/m_c = 0.21$.
so that the total mass of the $\{ud\}\hbox{-}\{ud\bar c\}$ diquark-triquark
is 
\begin{equation}
M_{\Theta_c} \approx 2778 + 207 = 2985 \ \hbox{MeV}\,.
\label{MThetac}
\end{equation}
Clearly, this is a rather rough estimate so is should be expected to hold
to no more than to within 50 MeV, so we expect 
$M_{\Theta_c} = 2985 \pm 50$ MeV.
In Ref.~\cite{NewPenta} the same approach gave an estimate 
$m_{\Theta^+}=1592$ MeV, overshooting the experimental value by about 
50 MeV,
so it is quite possible that the estimate \eqref{MThetac}
might be overshooting a bit as well, but at this point we don't think the
mass estimate is accurate enough to use the $\Theta^+$ input to fine-tune 
the prediction.

Assuming flavor $SU(3)$ symmetry, the general formula for the decay rate of a
baryon $B_1$ of mass $M_1$
into a baryon $B_2$ with mass $M_2$,
plus an octet pseudoscalar meson ${\cal P}$ of mass $M_{\cal P}$
is given by \cite{SMKNb}
\begin{equation}
\Gamma(B_1\to B_2+{\cal P}) = \frac{3 G_0^2}{2\pi (M_2+M_1)^2}|p|^3
\frac{M_2}{M_1}
\times {\cal C}
\label{theta_width}
\end{equation}
where $|\vec{p}|=\sqrt{(M_1^2-(M_2+M_{\cal P})^2)\cdot 
(M_1^2-(M_2-M_{\cal P})^2)}/2M_1$
is the momentum of the meson, $G_0$ is the appropriate coupling constant
(an analogue of $g_{\pi N N}$) and ${\cal C}$ is an $SU(3)_f$
 group-theoretical factor,
depending on the flavor and spin quantum numbers of the initial and final
state hadrons. Schematically, 
\beq{schematic_width}
\Gamma(B_1\to B_2+{\cal P}) =
G_0^2 \times {\cal C}
\times 
 \hbox{(phase space)}
\eeq
In \cite{SMKNb} eq.~\eqref{theta_width} was shown to compare well with
experiment for several well-measured decays of decuplet baryons, such as
$\Delta\rightarrow N \pi$,
$\Sigma^*\rightarrow \Lambda\pi$, etc., assuming $G_0\approx 19$,
with values of ${\cal C}$ which turned out to be between 1/15 and 1/5. 

Next, for
the decay $\Theta^+ \rightarrow K N$, it was assumed that $G_0\approx 9.5$
and ${\cal C}\approx 1/5$, yielding the prediction 
$\Gamma(\Theta^+ \rightarrow K N) = 15$ MeV, which seems to be in good
agreement with experiment \cite{Kyoto,Russia,Stepanyan:2003qr}.
On a qualitative level, the crucial observation is that the phase space for
$\Theta^+ \rightarrow K N$ is very small, so $\Theta^+$ is narrow, even 
though $G_0^2 \times {\cal C} \sim 20$.

In order to obtain a very rough estimate of the width of the decay 
$\Theta_c \rightarrow D N$, we use eq.~\eqref{theta_width}
with $M_1=M_{\Theta_c}=2985$ MeV, as given by \eqref{MThetac}, 
$M_2=M_N$, $M_{\cal P}=M_D$, $G_0\sim 10$ and ${\cal C}\sim 1/5$. This yields
\beq{GammaThetac}
\Gamma(\Theta_c \rightarrow D N) \ \sim \ 21\ \hbox{MeV}\,.
\eeq
Clearly, eq.~\eqref{GammaThetac} should be viewed only as indication of the
expected width, probably no better than a factor of 2. On the face of it,
this is for three reasons at least:
\hfill\break
(a) we do not know the value of the $g_{\pi N N}$ analogue, 
$g_{D\Theta_c N}=G_0$\,;
\hfill\break
(b) the group-theoretical factor $C$ was originally derived for
$SU(3)_f$\,;
\hfill\break
(c) the phase space is very sensitive to mass differences, so a relatively 
small shift in \thetac\ mass can cause a significant shift in its width.
\hfill\break
Still, the estimate \eqref{GammaThetac} is probably in the right ball park,
since $g_{D\Theta_c N}$ is a meson-baryon-baryon coupling, so assuming
$g_{D\Theta_c N}\sim 10$ is not unreasonable. As for the
group-theoretical factor ${\cal C}$, formally we can think of introducing 
(a very badly broken) flavor group $SU(3)$ encompassing the $u$, $d$ and
$c$ quarks. The group-theoretical factor reflects the group 
structure with no mass dependence, so changing $s$ to $c$ should not
affect it. 

On a qualitative level, it is important to realize that 
the phase space for $\Theta_c \rightarrow D N$ is small,
so for a typical hadronic coupling and a typical group-theoretical factor
we expect $\Theta_c$ to be narrow.

\subsection{The exotic baryon $\thetab=uudd\bar b$}

The above discussion can now be repeated,
this time replacing $\bar s$ by $\bar b$. We expect the \thetab\ to have the same
isospin and spin-parity as \thetac, i.e. 
an isosinglet with $J^P={1\over2}^-$\,. 

The \thetab\ mass is estimated exactly as for \thetac, i.e.
from eq.~\eqref{quarkmasses} we obtain
a very rough estimate of the $ud\bar b$ triquark effective mass,
$m_{ud\bar b}= 5770$ MeV, 
so that the reduced mass for the relative motion of the  
$ud$ diquark and $ud\bar b$ triquark system is 
$m_r(\{ud\}\hbox{-}\{ud\bar b\})= 640$ MeV.
As in the \thetac\ case, without the $P$-wave excitation energy
the $\{ud\}\hbox{-}\{ud\bar b\}$ diquark-triquark mass is
\begin{equation}
m^0_{\{ud\hbox{-}ud\bar b\}}= m_N + m_B
-{1\over 12}\,(1+\zeta_b)\, \left[M(\Delta)-M(N)\right] \approx \
6191\ \hbox{MeV}\,.
\label{M0uuddbbar}
\end{equation}
where $\zeta_b =  m_u/m_b = 0.07$,
so that the total mass of the $\{ud\}\hbox{-}\{ud\bar b\}$ diquark-triquark
is
\begin{equation}
M_{\thetab} \approx m^0_{\{ud\hbox{-}ud\bar b\}} + \delta E^{P-wave} =
6191 + 207 = 6398\pm 50\ \hbox{MeV}\,.
\label{MThetab}
\end{equation}
In order to obtain a very rough estimate of the width of the decay
$\thetab \rightarrow B N$, we again use eq.~\eqref{theta_width}
with $M_1=M_{\thetab}=6398$ MeV, as given by \eqref{MThetab},
$M_2=M_N$, $M_{\cal P}=M_B$, $G_0\sim 10$ and ${\cal C}\sim 1/5$. This yields
\beq{GammaThetab}
\Gamma(\thetab \rightarrow B N) \ \sim \ 4\ \hbox{MeV}\,.
\eeq
Again, we repeat here all the caveats regarding the very rough nature of this
estimate, as explained following 
the estimate of the \thetac\ width. On a qualitative level however, it seems
quite likely that the \thetac\ width will indeed be substantially more narrow
that the width of \thetac. 

\section*{Acknowledgments}

The research of one of us (M.K.) was supported in part by a grant from the
United States-Israel Binational Science Foundation (BSF), Jerusalem and
by the Einstein Center for Theoretical Physics at the Weizmann Institute.
The research of one of us (H.J.L.) was supported in part by the U.S. Department
of Energy, Division of High Energy Physics, Contract W-31-109-ENG-38.
We benefited from e-mail discussions with Ken Hicks and 
John Cumulat about experimental data.

%
\catcode`\@=11 
\def\references{ 
\ifpreprintsty \vskip 10ex
%
\hbox to\hsize{\hss \large \refname \hss }\else 
\vskip 24pt \hrule width\hsize \relax \vskip 1.6cm \fi \list 
{\@biblabel {\arabic {enumiv}}}
{\labelwidth \WidestRefLabelThusFar \labelsep 4pt \leftmargin \labelwidth 
\advance \leftmargin \labelsep \ifdim \baselinestretch pt>1 pt 
\parsep 4pt\relax \else \parsep 0pt\relax \fi \itemsep \parsep \usecounter 
{enumiv}\let \p@enumiv \@empty \def \theenumiv {\arabic {enumiv}}}
\let \newblock \relax \sloppy
 \clubpenalty 4000\widowpenalty 4000 \sfcode `\.=1000\relax \ifpreprintsty 
\else \small \fi}
\catcode`\@=12 
{\tighten

}

\end{document}